# Hippocampal and auditory contributions to speech segmentation

--Manuscript Draft--

| | |
|---|---|
| **Manuscript Number:** | CORTEX-D-21-00111R1 |
| **Article Type:** | Research Report |
| **Keywords:** | hippocampus; Statistical Learning; frequency tagging; sEEG; speech segmentation |
| **Corresponding Author:** | Clément François, Ph.D.<br>Aix-Marseille-University: Aix-Marseille Universite<br>Aix-en-Provence, FRANCE |
| **First Author:** | Neus Ramos-Escobar |
| **Order of Authors:** | Neus Ramos-Escobar |
| | Manuel Mercier |
| | Agnès Trébuchon-Fonséca |
| | Antoni Rodriguez-Fornells |
| | Clément François, Ph.D. |
| | Daniele Schön |
| **Abstract:** | Statistical learning has been proposed as a mechanism to structure and segment the continuous flow of information in several sensory modalities. Previous studies proposed that the medial temporal lobe, and in particular the hippocampus, may be crucial to parse the stream in the visual modality. However, the involvement of the hippocampus in auditory statistical learning, and specifically in speech segmentation is less clear. To explore the role of the hippocampus in speech segmentation based on statistical learning, we exposed seven pharmaco-resistant temporal lobe epilepsy patients to a continuous stream of trisyllabic pseudowords and recorded intracranial stereotaxic electro-encephalography (sEEG). We used frequency-tagging analysis to quantify neuronal synchronization of the hippocampus and auditory regions to the temporal structure of words and syllables of the stream. Results show that while auditory regions highly respond to syllable frequency, the hippocampus responds mostly to word frequency. These findings provide direct evidence of the involvement of the hippocampus in speech segmentation process and suggest a hierarchical organization of auditory information during speech processing. |
| **Response to Reviewers:** | Reviewer #1: Ramos-Escobar and colleagues presented continuous streams of auditory syllables to patients with intractable epilepsy followed by a forced choice recognition test on three-syllable words hidden in the streams. Using frequency tagging on intracranial EEG recordings on the surface of the auditory cortex and from depth electrodes in the medial temporal lobe, they measured statistical learning of syllables, two-syllables and three-syllable words. The authors report that the auditory cortex responds more to syllables than to words (i. e. shows a higher power in the frequency range at which syllables are presented than to the frequency at which words are presented), while the hippocampus responds more to words than to syllables (shows a higher power in the frequency range at which words are presented than to the frequency at which syllables are presented). Based on those findings the authors conclude that statistical learning is hierarchically organized in the brain,<br>and that the hippocampus plays an important role in statistical learning of speech.<br><br>The manuscript is well written and clear and shows interesting and compelling results.<br><br>I have a few comments which need attention before I can recommend publication.<br>Major comment:<br>-   The most puzzling issue is the finding that behavior shows a strong discrepancy to the neural responses. The authors mention that the force choice has low sensitivity to learning (page 11, line13). This may be true, but alternatively, the hippocampus may not be necessary for statistical learning (as was previously mentioned in line 1 of the same page). It would be good to discuss that possibility too in the context of learning |

speech sequences and in the context of the damage of the MTL in these patients. It would also be good to suggest alternative behavioral methods which would reveal learning.

We thank the reviewer for this comment. We have further developed this point in the discussion and also refer to other studies that have shown similar discrepancies (e.g. Henin et al., 2021). Most importantly, in relation to this point, we now report the analysis of sEEG data acquired during the test phase together with the corresponding figure (Figure 4). These results show that the event-related potentials (ERPs) to words and nonwords differ in the hippocampus. In other words, neurophysiological data show that 1) the hippocampus contributes to speech stream segmentation, as seen during the learning phase, 2) the hippocampus is sensitive to the familiarity of the items during the test phase (thus in a different dataset). Then, the absence of behavioural learning seems to be due to a high noise at the decision making level. We now further discuss this point in relation to possible weaknesses of the behavioural task and make reference to newly developed experimental designs (François et al., 2016, 2017). Finally, we would like to clarify that we do not make any causal statement in the manuscript and that our data only show that the hippocampus is involved in speech segmentation, but not that it is necessary for speech segmentation as these claims possibly require perturbation or lesion studies.

Results section:

Page 10, lines 235-241: "Importantly, however, as shown on Figure 4, the ERP data show a significant difference between words and nonwords in hippocampal channels in the 250-400 (beta = -18.8; CI = -33.3 -4.2; p <.01) and 550-700 ms (beta = -19.6, CI = -35.9 -3.2; p < .01) time-windows. A significant effect over a single 50ms time window, between 350 and 400 ms, is also found over auditory channels (beta = -8.4, CI = -16.5 -0.7; p < .05). Overall, these results confirm that patients did segment the words during the learning phase and that the hippocampus is particularly sensitive to the familiarity of the items."
Discussion section:
Page 12-13, lines 289-327: "In the current work, patients, most of whom had temporal lobe epilepsy, performed poorly in the explicit recognition test as patients with MTL lesions. By contrast, they presented robust neural tuning at target frequencies corresponding to different levels of the speech hierarchy (i.e., word, syllable, and pair of syllables) during the learning phase. This result indicates that learning did take place and that the hippocampus was functional with respect to statistical learning. It also confirms that implicit online measures of learning based on electrophysiological data are more sensitive than behavioural measures (François, Tillmann & Schön, 2012). Indeed, the analysis of the ERPs collected during the 2AFC task also revealed significant differences between words and nonwords over hippocampal channels. This result fits well with previous studies on speech segmentation based on SL showing functional activations of the hippocampus during speech segmentation tasks (Turk-Browne et al., 2009; Schapiro, Kustner, & Turk-Browne 2012; Schapiro et al., 2016; Barascud et al., 2016). A similar familiarity effect has been also reported when focusing on the 2AFC test (François & Schön, 2010, 2011; De Diego Balaguer et al., 2007). These studies used scalp EEG to show that healthy adults exhibited a larger negativity for unfamiliar than for newly learned. However, the percentage of correct explicit word recognition did not differ from chance level. Similar discrepancies between behavioural and neural data have been reported in previous neuroimaging studies of speech segmentation based on SL in healthy adults (François & Schön, 2010, 2011; McNealy et al., 2006; Turk-Browne et al., 2009; Sanders et al., 2002) and in patients with MTL damage (Henin et al., 2021; Schapiro et al., 2014; Covington, Brown-Schmidt & Duff, 2018). Moreover, the role of the hippocampus and MTL region during recognition memory tasks has largely been demonstrated in both healthy adults and patients with damage to the MTL (Brown & Aggleton, 2001; Düzel et al., 2001; Eldridge et al., 2000; Stark & Squire, 2000; Ranganath et al., 2004). Here, we used an implicit procedure during the learning phase and evaluated the learning using an explicit behavioural task that requires the conscious recognition of word-forms presented auditorily. While our approach has the advantage of being of a very short duration, the 2AFC task has been largely criticized for its low sensitivity due to different factors (François, Tillmann & Schön, 2012; Batterink et al., 2015; Siegelman, Bogaerts & Frost, 2017; Siegelman et al., 2018; Frost, Armstrong & Christiansen, 2019; Christiansen, 2019; ). For instance, the AFC task requires participants to make an explicit judgment on two presented items without feedback, which might be particularly challenging in the case of the relatively weak memory traces created during the implicit learning phase (Schön & François, 2011; Rodriguez-Fornells et al., 2009). Moreover, the design of the AFC test trials does not allow differentiating between word recognition and nonword rejection as it is the case when using a lexical decision task (François et al., 2016; Ramos-Escobar et al., 2021). Recent studies on speech segmentation based on SL have elegantly proposed innovative designs to overcome the weaknesses associated with the use of explicit tests. Of particular relevance is the use of implicit measures such as EEG, sEEG, or Reaction-Times collected during the learning or an online test phase (see for example François et al., 2016, 2017; de Diego Balaguer et al., 2007 for the analysis of ERPs to illegal items without explicit recognition) that seem more appropriate and sensitive to fully capture implicit learning processes (Kim, Seitz, Feenstra, & Shams, 2009; Kóbor et al., 2020; Turk-Browne et al., 2005; Batterink & Paller, 2017; Siegelman, Bogaerts & Frost, 2017)."

Minor comments:

- The experimental methods are very briefly described and it is difficult to understand the flow of events, particularly the duration of the trial or trials and the test phase. It would help to move the sentence from the stimuli section "Each word is presented 60 times …" up to the experimental procedure section. Without the understanding that there is only one stream it is also difficult to understand the segmenting of the EEG signal.

We thank the reviewer for this comment. We agree that the experimental method should be developed further to facilitate the replication of the study. Therefore, we have added more details in the method section. We now also acknowledge that the procedure that we used here was similar to the one used in various studies of our group with healthy adults and children (François & Schön, 2010; 2011; François et al., 2013; 2014).

Page 6-7, lines 152-167: "We used a similar experimental design to the one used in our previous studies with healthy adults and children (Schön et al., 2008; François & Schön 2010; 2011; François et al., 2013; 2014). Specifically, the experimental procedure consisted of two consecutive phases, an implicit learning phase followed by an explicit 2-alternative forced-choice (2AFC) task. Before starting the implicit learning phase, patients were asked to listen carefully to one single auditory stream without explicit instructions of learning (see Stimuli section for a description of the speech streams). Importantly, we did our best to keep the entire procedure implicit. During the learning phase, patients were exposed to a single continuous speech stream that was composed of 4 pseudo-words presented 60 times each, thus leading to a single continuous stream of 240 words that lasted 4 min. Immediately after this learning phase, patients performed the behavioural 2AFC task that lasted 5 min. During each trial of the test, patients were presented with two consecutive auditory words and had to press one of two buttons to indicate which of two words (first or second item) most closely resembled what they had just heard in the continuous stream (see Figure 2). Importantly, one test item was a word from the learning stream while the other was a "nonword" that was never heard before the test. Each familiar word of the language (word) was presented with each unfamiliar word (nonwords), making up 16 pairs that were repeated twice, thus leading to 32 test trials."
"

- The authors mention that "epochs time-locked at the onset of each word were created by segmenting the recordings from 4 words before and 4 after the stimulus yielding epochs of 8-word length (lasting 7.2s)." I don't understand that sentence. Shouldn't 4+4+1 be 9 word length? Or is the word itself included in the "4 after the stimulus"?

We apologize for this misunderstanding. The epoch is defined with respect to the word onset, so it consists of 4 words before and four words after the onset. We have rephrased this sentence.
Page 8, lines 190-192: "Then, epochs time-locked to the onset of each word were created by segmenting the continuous EEG data from 4 words before and 4 after the stimulus yielding epochs of 8-word length (lasting 7.2s)."

- The syntax in the legend of Figure 3: "Black arrows indicate the bar where falls …" should be corrected

We thank the reviewer for pointing this out. We have rephrased the legend as following:
“Black arrows indicate the bin where the hippocampal power response falls.”

- Delete ; at the end of the citations on page 11 in line 16.

This has been done.

Reviewer #2: The authors provide an interesting examination of statistical learning using intracranial recordings in patients with epilepsy. Specifically, using frequency-tagged auditory stimuli they reported observing greater entrained responses in the hippocampus to artificial words and greater entrained responses in auditory cortex to phonemes. Studies with intracranial recordings (sEEG/ECoG) remain uncommon and valuable datasets for human neuroscience research. At present, however, the strength of the results is unclear to this reader, expanded on below, as it is possible that the pattern of results observed is unrelated to statistical learning, instead reflecting the particular set of analyses employed.

Primary Concerns

Statistical comparisons. The study examined significance within subjects by comparing power across electrodes. This is less commonly used than comparing power at each individual electrode to some baseline -- given the continuous nature of the stimulus in this experiment, I would expect the use of a prestimulus resting state period. The major weakness of the current approach is that non-baselined power will reflect a mixture of intrinsic power and evoked power, particularly because there was no temporal jitter between presentation of the 240 stimuli. Moreover, this measure of relative power across electrodes is dependent on where the other electrodes are located -- if a patient had auditory and hippocampal electrodes that each responded strongly to phonemes, then neither would be significant.

We understand the reviewer’s concern. The choice of comparing power across electrodes was constrained by the absence of a sufficiently long baseline. Indeed, ideally, one would need a baseline as long as the learning phase in order to have an equivalent SNR. This was clearly not the case due to clinical constraints requiring to keep the experiment as short as possible.
However, we would like to argue that our approach is actually more conservative than testing against a baseline. Indeed entrainment to auditory stimuli will be apparent in many regions (not limited to the auditory cortex, see Pesnot et al., 2021). Thus, thresholding using the distribution of the whole dataset is more conservative than using the baseline that will present NO entrainment but only intrinsic oscillatory activity. What we now tried to clarify, and that is important in this context, is the fact that by computing averages, we remove non time-locked activity (intrinsic oscillations) and only focus on evoked activity.

Page 8, lines 196-199: “Importantly, by computing averages, similarly to other frequency tagging studies (Nozaradan et al., 2021; Jonas et al., 2016), we remove non time-locked activity (intrinsic oscillations), enhance the signal-to-noise ratio of EEG activities time locked to the patterns and only focus on evoked activity.”

Below, we computed the same power analysis on a surrogate data built by randomly picking non time-locked epochs for one patient. Such a surrogate distribution, possibly simulating a baseline thresholding strategy, shows extremely low values at the frequencies of interest compared to the real data (top panel). This shows that our approach is possibly more conservative than using a baseline approach: the probability of one single value (e.g., in the hippocampus) being above threshold by chance is smaller.

The reviewer has another related remark: whether this measure of relative power across electrodes is dependent on where the other electrodes are located; “if a patient had auditory and hippocampal electrodes that each responded strongly to phonemes, then neither would be significant”. This is indeed correct, BUT we do systematically have many more contacts in regions outside the auditory and hippocampal areas than inside these areas. Patients have between ~140 and ~200 useful contacts and only a few of these (<10) are located in the hippocampus and auditory regions (<10).

Page 9, lines 209-211: “For each patient and for each target frequency (word, syllable & two syllables), we computed the distribution of power values across all contacts (between 140 and 200 per patient, spanning several brain regions beyond the primary auditory cortex and the hippocampus).”

Statistical learning. Patients did not demonstrate behavioral effects of statistical learning, and so it's possible that they were unaware which syllable groups formed word boundaries. It appears the test phase data was not analyzed, which could lend credibility to the authors' claim that subjects implicitly learned the statistical representation. More generally, 4 minutes of a stimulus may be too short a period for learning to occur in these patients. If the authors split their data in half, can they show that frequency-tagged responses to words increased whereas other syllable frequency stayed the same?

We thank the reviewer for this comment. We now report the neurophysiological data acquired during the behavioural task, namely the testing phase following the learning phase. These results show that the ERPs to words and pseudowords differ in the hippocampus. In other words, neurophysiological data show that 1) the hippocampus contributes to speech stream segmentation, as seen during the learning phase, 2) the hippocampus is sensitive to the familiarity of the items during the test phase (thus in a different dataset). Then, the absence of behavioural learning seems to be due to a high noise at the decision making level. We now discuss this point also in relation to some weaknesses of the behavioural task.

Page 12-13, lines 289-327: “In the current work, patients, most of whom had temporal lobe epilepsy, performed poorly in the explicit recognition test as patients with MTL lesions. By contrast, they presented robust neural tuning at target frequencies corresponding to different levels of the speech hierarchy (i.e., word, syllable, and pair of syllables) during the learning phase. This result indicates that learning did take place and that the hippocampus was functional with respect to statistical learning. It also confirms that implicit online measures of learning based on electrophysiological data are more sensitive than behavioural measures (François, Tillmann & Schön, 2012). Indeed, the analysis of the ERPs collected during the 2AFC task also revealed significant differences between words and nonwords over hippocampal channels. This result fits well with previous studies on speech segmentation based on SL showing functional activations of the hippocampus during speech segmentation tasks (Turk-Browne et al., 2009; Schapiro, Kustner, & Turk-Browne 2012; Schapiro et al., 2016; Barascud et al., 2016). A similar familiarity effect has been also reported when focusing on the 2AFC test (François & Schön, 2010, 2011; De Diego Balaguer et al., 2007). These studies used scalp EEG to show that healthy adults exhibited a larger negativity for unfamiliar than for newly learned. However, the percentage of correct explicit word recognition did not differ from chance level. Similar discrepancies between behavioural and neural data have been reported in previous neuroimaging studies of speech segmentation based on SL in healthy adults (François & Schön, 2010, 2011; McNealy et al., 2006; Turk-Browne et al., 2009; Sanders et al., 2002) and in patients with MTL damage (Henin et al., 2021; Schapiro et al., 2014; Covington, Brown-Schmidt & Duff, 2018). Moreover, the role of the hippocampus and MTL region during recognition memory tasks has largely been demonstrated in both healthy adults and patients with damage to the MTL (Brown & Aggleton, 2001; Düzel et al., 2001; Eldridge et al., 2000; Stark & Squire, 2000; Ranganath et al., 2004). Here, we used an implicit procedure during the learning phase and evaluated the learning using an explicit behavioural task that requires the conscious recognition of word-forms presented auditorily. While our approach has the advantage of being of a very short duration, the 2AFC task has been largely criticized for its low sensitivity due to different factors (François, Tillmann & Schön, 2012; Batterink et al., 2015; Siegelman, Bogaerts & Frost, 2017; Siegelman et al., 2018; Frost, Armstrong & Christiansen, 2019; Christiansen, 2019; ). For instance, the AFC task requires participants to make an explicit judgment on two presented items without feedback, which might be particularly challenging in the case of the relatively weak memory traces created during the implicit learning phase (Schön & François, 2011; Rodriguez-Fornells et al., 2009). Moreover, the design of the AFC test trials does not allow differentiating between word recognition and nonword rejection as it is the case when using a lexical decision task (François et al., 2016; Ramos-Escobar et al., 2021). Recent studies on speech segmentation based on SL have elegantly proposed innovative designs to overcome the weaknesses associated with the use of explicit tests. Of particular relevance is the use of implicit measures such as EEG, sEEG, or Reaction-Times collected during the learning or an online test phase (see for example François et al., 2016, 2017; de Diego Balaguer et al., 2007 for the analysis of ERPs to illegal items without explicit recognition) that seem more appropriate and sensitive to fully capture implicit learning processes (Kim, Seitz, Feenstra, & Shams, 2009; Kóbor et al., 2020; Turk-Browne et al., 2005; Batterink & Paller, 2017; Siegelman, Bogaerts & Frost, 2017)."

Concerning the possibility of splitting the data, we followed the reviewer suggestion. However, as the reviewer can see in the figure below, the effect is not clear cut, although there is a tendency for an increase at the word frequency. This is possibily due to different learning curves in the different patients that may prevent observing a clear increase. We also tried to have a more temporally resolved analysis to explore inter-individual differences, but the estimate became too noisy when using small data sets (e.g. 8 periods of 30 seconds). We eventually decided not to report this analysis in the manuscript.

Secondary Comments

4 ms is a very short baseline period which can introduce noise to the analysis. Do the authors have justification over a longer baseline (at least 100 ms)?

Sorry this was a typo error, it should be seconds and correspond to half of the window.

The authors mention normalization in the methods. How was power normalized?
A common approach with frequency-tagging is to replot the data as signal-to-noise ratios, wherein power at the target frequency is compared against neighboring frequencies to cancel out the effects of the 1/f distribution.

We agree with the reviewer that some studies have used such a normalization procedure. However, we think that in the case of sEEG recordings the SNR is much higher than with scalp data. The suggested procedure that implicitly increases the local SNR may not be necessary in our case and we prefer not to use it and to show the 'true' FFT. Please also note that, as detailed above, we do not have the 1/f in the PSD because we work on averages. Further, recent studies have used similar approaches to study the neural mechanisms supporting the extraction of speech units based on SL in adults and children (see Jonas et al., 2016; Ordin et al., 2020; Ramos-Escobar et al., 2021).

Why was evoked power calculated as opposed to total power averaged across the entire time-range? Evoked power, when no jitter across trials, can lead to peaks at intrinsic oscillations. Moreover, total power would enable a plot of the 1/f distributions for electrodes and subjects which can be helpful in evaluating the quality of the recordings.

As we clarified above, the strategy of averaging is commonly used (see for instance Nozaradan et al., 2021; Jonas et al., 2016) in frequency tagging analysis to enhance the signal-to-noise ratio of EEG activities time locked to the patterns. Below, we computed the full range power spectral density for each patient (colored lines) for both hippocampal (top) and auditory (bottom) channels. On the left, the reviewer can appreciate that it is not easy to see much on the regular PSD of hippocampal channels. The scenario becomes a little bit better when normalizing by neighbours (dividing each value by two neighbour values), as can be seen on the right part of the figure. However, while for the auditory cortex, that has a very strong response to the syllabic

| | |
|---|---|
| | rate, the result is clear cut, for the hippocampal channels, have smaller responses, results are less clear and mostly visible in the first harmonic of the word frequency (2.2 Hz). We feel that this well illustrates the advantage of computing the FFT of a sliding average. Also, note that, as reported in the methods section, we cautiously use an overlap equal to twice the size of the word duration to ensure that possible artifacts would not lead to a spurious peak at the word frequency.<br><br>Assuming the power effects are driven by the stimuli, is it possible that the hippocampus tracked 'words' because the task required discrimination of 3 phoneme groups? Were subjects aware what they would be tested on?<br><br>We thank the reviewer for this comment. In this specific case, the answer is no. We used an implicit version of the SL paradigm in which the patients were not aware of the purpose of the task nor that they would be tested afterward. We agree that some studies have used explicit instructions of learning which may have triggered different cognitive mechanisms (Cunillera et al., 2006, 2009). Again, here, the patients were only instructed to listen carefully to an auditory stream without explicit instructions of learning. Importantly, the grouping of phonemes can only be done by statistical learning as there are no other (e.g., acoustic) cues to group the individual phonemes. |

Title Page

**Hippocampal and auditory contributions to speech segmentation**

Neus Ramos-Escobar[a,b] , Manuel Mercier[c], Agnès Trébuchon-Fonséca[c,d], Antoni Rodriguez-Fornells[a,b,e], Clément François[f]*, Daniele Schön[c]*

[a]Dept. of Cognition, Development and Educational Science, Institute of Neuroscience, University of Barcelona, L'Hospitalet de Llobregat, Barcelona, 08097, Spain.

[b]Cognition and Brain Plasticity Group, Bellvitge Biomedical Research Institute (IDIBELL), L'Hospitalet de Llobregat, Barcelona, 08097, Spain

[c]Aix Marseille Univ, Inserm, INS, Inst Neurosci Syst, Marseille, France

[d]APHM, Hôpital de la Timone, Service de Neurophysiologie Clinique, Marseille, France

[e]Catalan Institution for Research and Advanced Studies, ICREA, Barcelona, Spain

[f]Aix Marseille Univ, CNRS, LPL, (13100) Aix-en-Provence, France

* co-senior authorship

*Corresponding Authors: Daniele Schön: daniele.schon@univ-amu.fr +33 491324100 and Clément François: clement.françois@univ-amu.fr +33 413552714

**Error!Error!Error!Error!**

Marseille, 2nd of November 2021

Dear Editor,

Thank you for giving us with the opportunity to submit a revised version of our work. Please find attached the revised version of our manuscript entitled "*Hippocampal and auditory contributions to speech segmentation*", which we would like considered for publication in *Cortex*.

We are grateful to the two reviewers for all their helpful comments and interesting suggestions. We feel that we were able to address all the suggestions in an appropriate manner. You will find our detailed answers in the "responses to the reviewers" section but we would like to acknowledge some specific points that have been raised during the review.

Both reviewers had concern about the experimental procedure, the methods and the analyses we used. Therefore, we have provided further details about each of these points and have added new results with the corresponding figure in the new version of the manuscript. Based on the reviewers' comments, we have added new analyses focusing on the ERPs of the 2AFC test and discuss these new results in the discussion section. However, we have preferred not include the results comparing the two halves of the learning phase nor those obtained with the neighboring normalization. We will be delighted to add them in a new version of the manuscript if the editor considers these results important.

Thank you very much in advance for your consideration
Sincerely yours,

Neus Ramos-Escobar, Manuel Mercier, Agnés Trébuchon, Antoni Rodriguez-Fornells, Clément François & Daniele Schön

**Reviewer #1:** Ramos-Escobar and colleagues presented continuous streams of auditory syllables to patients with intractable epilepsy followed by a forced choice recognition test on three-syllable words hidden in the streams. Using frequency tagging on intracranial EEG recordings on the surface of the auditory cortex and from depth electrodes in the medial temporal lobe, they measured statistical learning of syllables, two-syllables and three-syllable words. The authors report that the auditory cortex responds more to syllables than to words (i. e. shows a higher power in the frequency range at which syllables are presented than to the frequency at which words are presented), while the hippocampus responds more to words than to syllables (shows a higher power in the frequency range at which words are presented than to the frequency at which syllables are presented). Based on those findings the authors conclude that statistical learning is hierarchically organized in the brain,
and that the hippocampus plays an important role in statistical learning of speech.

The manuscript is well written and clear and shows interesting and compelling results.

I have a few comments which need attention before I can recommend publication.
Major comment:
- The most puzzling issue is the finding that behavior shows a strong discrepancy to the neural responses. The authors mention that the force choice has low sensitivity to learning (page 11, line13). This may be true, but alternatively, the hippocampus may not be necessary for statistical learning (as was previously mentioned in line 1 of the same page). It would be good to discuss that possibility too in the context of learning speech sequences and in the context of the damage of the MTL in these patients. It would also be good to suggest alternative behavioral methods which would reveal learning.

We thank the reviewer for this comment. We have further developed this point in the discussion and also refer to other studies that have shown similar discrepancies (e.g. Henin et al., 2021). Most importantly, in relation to this point, we now report the analysis of sEEG data acquired during the test phase together with the corresponding figure (Figure 4). These results show that the event-related potentials (ERPs) to words and nonwords differ in the hippocampus. In other words, neurophysiological data show that 1) the hippocampus contributes to speech stream segmentation, as seen during the learning phase, 2) the hippocampus is sensitive to the familiarity of the items during the test phase (thus in a different dataset). Then, the absence of behavioural learning seems to be due to a high noise at the decision making level. We now further discuss this point in relation to possible weaknesses of the behavioural task and make reference to newly developed experimental designs (François et al., 2016, 2017). Finally, we would like to clarify that we do not make any causal statement in the manuscript and that our data only show that the hippocampus is involved in speech segmentation, but not that it is necessary for speech segmentation as these claims possibly require perturbation or lesion studies.

Results section:

Page 10, lines 235-241: ”*Importantly, however, as shown on **Figure 4**, the ERP data show a significant difference between words and nonwords in hippocampal channels in the 250-400 (beta = -18.8; CI = -33.3 -4.2; p <.01) and 550-700 ms (beta = -19.6, CI = -35.9 -3.2; p < .01) time-windows. A significant effect over a single 50ms time window, between 350 and 400 ms, is also found over auditory channels (beta = -8.4, CI = -16.5 -0.7; p < .05). Overall, these results confirm that patients* *did segment the words during the learning phase and that the hippocampus is particularly sensitive to the familiarity of the items.*"

Discussion section:

Page 12-13, lines 289-327: "*In the current work, patients, most of whom had temporal lobe epilepsy, performed poorly in the explicit recognition test as patients with MTL lesions. By contrast, they presented robust neural tuning at target frequencies corresponding to different levels of the speech hierarchy (i.e., word, syllable, and pair of syllables) during the learning phase. This result indicates that learning did take place and that the hippocampus was functional with respect to statistical learning. It also confirms that implicit online measures of learning based on electrophysiological data are more sensitive than behavioural measures (François, Tillmann & Schön, 2012). Indeed, the analysis of the ERPs collected during the 2AFC task also revealed significant differences between words and nonwords over hippocampal channels. This result fits well with previous studies on speech segmentation based on SL showing functional activations of the hippocampus during speech segmentation tasks (Turk-Browne et al., 2009; Schapiro, Kustner, & Turk-Browne 2012; Schapiro et al., 2016; Barascud et al., 2016). A similar familiarity effect has been also reported when focusing on the 2AFC test (François & Schön, 2010, 2011; De Diego Balaguer et al., 2007). These studies used scalp EEG to show that healthy adults exhibited a larger negativity for unfamiliar than for newly learned. However, the percentage of correct explicit word recognition did not differ from chance level. Similar discrepancies between behavioural and neural data have been reported in previous neuroimaging studies of speech segmentation based on SL in healthy adults (François & Schön, 2010, 2011; McNealy et al., 2006; Turk-Browne et al., 2009; Sanders et al., 2002) and in patients with MTL damage (Henin et al., 2021; Schapiro et al., 2014; Covington, Brown-Schmidt & Duff, 2018). Moreover, the role of the hippocampus and MTL region during recognition memory tasks has largely been demonstrated in both healthy adults and patients with damage to the MTL (Brown & Aggleton, 2001; Düzel et al., 2001; Eldridge et al., 2000; Stark & Squire, 2000; Ranganath et al., 2004). Here, we used an implicit procedure during the learning phase and evaluated the learning using an explicit behavioural task that requires the conscious recognition of word-forms presented auditorily. While our approach has the advantage of being of a very short duration, the 2AFC task has been largely criticized for its low sensitivity due to different factors (François, Tillmann & Schön, 2012; Batterink et al., 2015; Siegelman, Bogaerts & Frost, 2017; Siegelman et al., 2018; Frost, Armstrong & Christiansen, 2019; Christiansen, 2019; ). For instance, the AFC task requires participants to make an explicit judgment on two presented items without feedback, which might be particularly challenging in the case of the relatively weak memory traces created during the implicit learning phase (Schön & François, 2011; Rodriguez-Fornells et al., 2009). Moreover, the design of the AFC test trials does not allow differentiating between word recognition and nonword rejection as it is the case when using a lexical decision task (François et al., 2016; Ramos-Escobar et al., 2021). Recent studies on speech* *segmentation based on SL have elegantly proposed innovative designs to overcome the weaknesses associated with the use of explicit tests. Of particular relevance is the use of implicit measures such as EEG, sEEG, or Reaction-Times collected during the learning or an online test phase (see for example François et al., 2016, 2017; de Diego Balaguer et al., 2007 for the analysis of ERPs to illegal items without explicit recognition) that seem more appropriate and sensitive to fully capture implicit learning processes (Kim, Seitz, Feenstra, & Shams, 2009; Kóbor et al., 2020; Turk-Browne et al., 2005; Batterink & Paller, 2017; Siegelman, Bogaerts & Frost, 2017).*"

Minor comments:

- The experimental methods are very briefly described and it is difficult to understand the flow of events, particularly the duration of the trial or trials and the test phase. It would help to move the sentence from the stimuli section "Each word is presented 60 times …" up to the experimental procedure section. Without the understanding that there is only one stream it is also difficult to understand the segmenting of the EEG signal.

We thank the reviewer for this comment. We agree that the experimental method should be developed further to facilitate the replication of the study. Therefore, we have added more details in the method section. We now also acknowledge that the procedure that we used here was similar to the one used in various studies of our group with healthy adults and children (François & Schön, 2010; 2011; François et al., 2013; 2014).

Page 6-7, lines 152-167: *"We used a similar experimental design to the one used in our previous studies with healthy adults and children (Schön et al., 2008; François & Schön 2010; 2011; François et al., 2013; 2014). Specifically, the experimental procedure consisted of two consecutive phases, an implicit learning phase followed by an explicit 2-alternative forced-choice (2AFC) task. Before starting the implicit learning phase, patients were asked to listen carefully to one single auditory stream without explicit instructions of learning (see Stimuli section for a description of the speech streams). Importantly, we did our best to keep the entire procedure implicit. During the learning phase, patients were exposed to a single continuous speech stream that was composed of 4 pseudo-words presented 60 times each, thus leading to a single continuous stream of 240 words that lasted 4 min. Immediately after this learning phase, patients performed the behavioural 2AFC task that lasted 5 min. During each trial of the test, patients were presented with two consecutive auditory words and had to press one of two buttons to indicate which of two words (first or second item) most closely resembled what they had just heard in the continuous stream (see Figure 2). Importantly, one test item was a word from the learning stream while the other was a "nonword" that was never heard before the test. Each familiar word of the language (word) was presented with each unfamiliar word (nonwords), making up 16 pairs that were repeated twice, thus leading to 32 test trials.*"

"

- The authors mention that "epochs time-locked at the onset of each word were created by segmenting the recordings from 4 words before and 4 after the stimulus yielding epochs of 8-word length (lasting 7.2s)." I don't understand that sentence. Shouldn't 4+4+1 be 9 word length? Or is the word itself included in the "4 after the stimulus"?

We apologize for this misunderstanding. The epoch is defined with respect to the word onset, so it consists of 4 words before and four words after the onset. We have rephrased this sentence.

Page 8, lines 190-192: *"Then, epochs time-locked to the onset of each word were created by segmenting the continuous EEG data from 4 words before and 4 after the stimulus yielding epochs of 8-word length (lasting 7.2s)."*

- The syntax in the legend of Figure 3: "Black arrows indicate the bar where falls …" should be corrected

We thank the reviewer for pointing this out. We have rephrased the legend as following:
*"Black arrows indicate the bin where the hippocampal power response falls."*

- Delete ; at the end of the citations on page 11 in line 16.

This has been done.

**Reviewer #2**: The authors provide an interesting examination of statistical learning using intracranial recordings in patients with epilepsy. Specifically, using frequency-tagged auditory stimuli they reported observing greater entrained responses in the hippocampus to artificial words and greater entrained responses in auditory cortex to phonemes. Studies with intracranial recordings (sEEG/ECoG) remain uncommon and valuable datasets for human neuroscience research. At present, however, the strength of the results is unclear to this reader, expanded on below, as it is possible that the pattern of results observed is unrelated to statistical learning, instead reflecting the particular set of analyses employed.

Primary Concerns

Statistical comparisons. The study examined significance within subjects by comparing power across electrodes. This is less commonly used than comparing power at each individual electrode to some baseline -- given the continuous nature of the stimulus in this experiment, I would expect the use of a prestimulus resting state period. The major weakness of the current approach is that non-baselined power will reflect a mixture of intrinsic power and evoked power, particularly because there was no temporal jitter between presentation of the 240 stimuli. Moreover, this measure of relative power across electrodes is dependent on where the other electrodes are located -- if a patient had auditory and hippocampal electrodes that each responded strongly to phonemes, then neither would be significant.

We understand the reviewer's concern. The choice of comparing power across electrodes was constrained by the absence of a sufficiently long baseline. Indeed, ideally, one would need a baseline as long as the learning phase in order to have an equivalent SNR. This was clearly not the case due to clinical constraints requiring to keep the experiment as short as possible. However, we would like to argue that our approach is actually more conservative than testing against a baseline. Indeed entrainment to auditory stimuli will be apparent in many regions (not limited to the auditory cortex, see Pesnot et al., 2021). Thus, thresholding using the distribution of the whole dataset is more conservative than using the baseline that will present NO entrainment but only intrinsic oscillatory activity. What we now tried to clarify, and that is important in this context, is the fact that by computing averages, we remove non time-locked activity (intrinsic oscillations) and only focus on evoked activity.

Page 8, lines 196-199: *"Importantly, by computing averages, similarly to other frequency tagging studies (Nozaradan et al., 2021; Jonas et al., 2016), we remove non time-locked activity (intrinsic oscillations), enhance the signal-to-noise ratio of EEG activities time locked to the patterns and only focus on evoked activity."*

Below, we computed the same power analysis on a surrogate data built by randomly picking non time-locked epochs for one patient. Such a surrogate distribution, possibly simulating a baseline thresholding strategy, shows extremely low values at the frequencies of interest compared to the real data (top panel). This shows that our approach is possibly more conservative than using a baseline approach: the probability of one single value (e.g., in the hippocampus) being above threshold by chance is smaller.

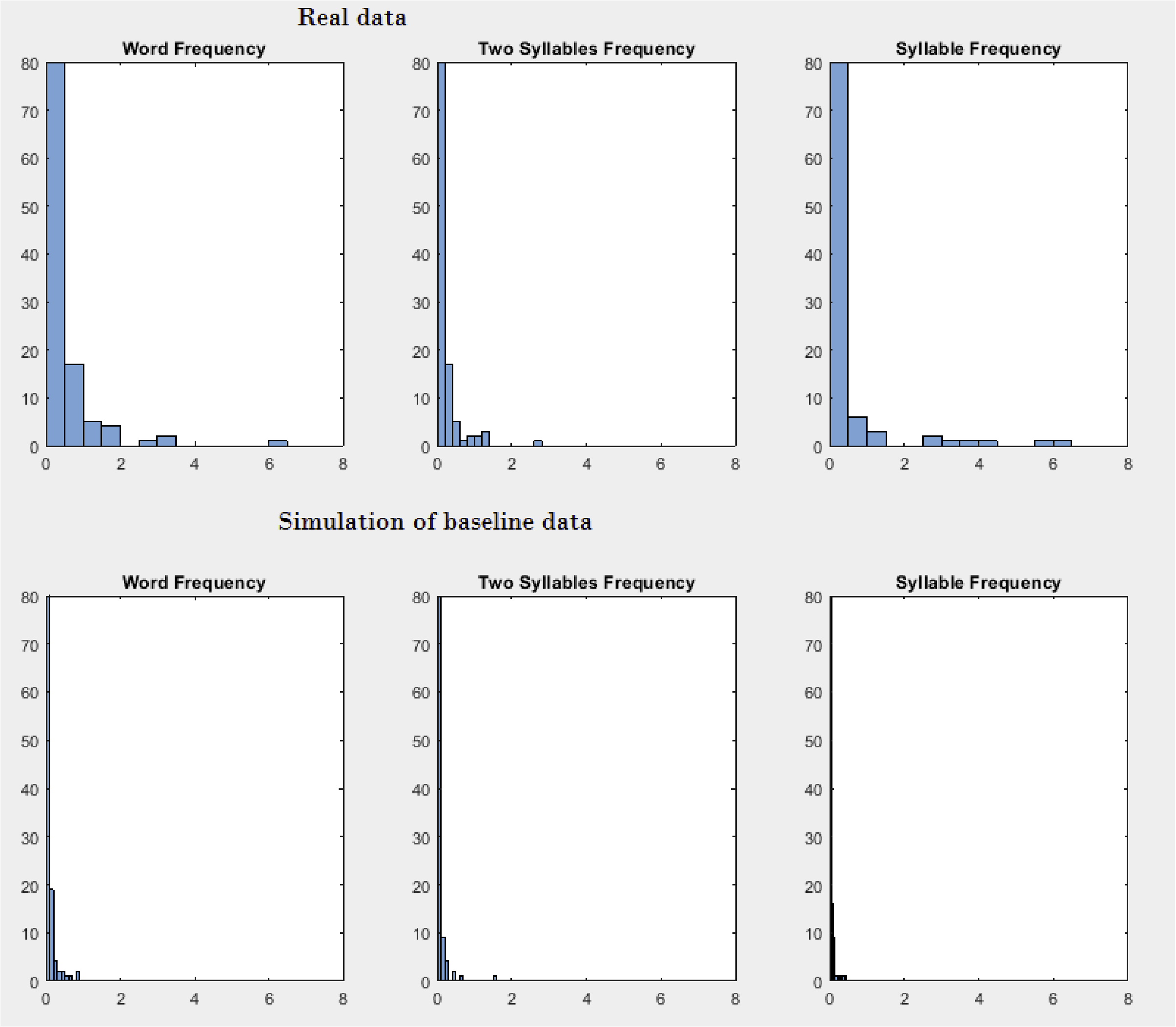

The reviewer has another related remark: whether this measure of relative power across electrodes is dependent on where the other electrodes are located; "if a patient had auditory and hippocampal electrodes that each responded strongly to phonemes, then neither would be significant". This is indeed correct, BUT we do systematically have many more contacts in regions outside the auditory and hippocampal areas than inside these areas. Patients have between ~140 and ~200 useful contacts and only a few of these (<10) are located in the hippocampus and auditory regions (<10).

Page 9, lines 209-211: "*For each patient and for each target frequency (word, syllable & two syllables), we computed the distribution of power values across all contacts (between 140 and 200 per patient, spanning several brain regions beyond the primary auditory cortex and the hippocampus).*"

Statistical learning. Patients did not demonstrate behavioral effects of statistical learning, and so it's possible that they were unaware which syllable groups formed word boundaries. It appears the test phase data was not analyzed, which could lend credibility to the authors' claim that subjects implicitly learned the statistical representation. More generally, 4 minutes of a stimulus may be too short a period for learning to occur in these patients. If the authors split their data in half, can they show that frequency-tagged responses to words increased whereas other syllable frequency stayed the same?

We thank the reviewer for this comment. We now report the neurophysiological data acquired during the behavioural task, namely the testing phase following the learning phase. These results show that the ERPs to words and pseudowords differ in the hippocampus. In other words, neurophysiological data show that 1) the hippocampus contributes to speech stream segmentation, as seen during the learning phase, 2) the hippocampus is sensitive to the familiarity of the items during the test phase (thus in a different dataset). Then, the absence of behavioural learning seems to be due to a high noise at the decision making level. We now discuss this point also in relation to some weaknesses of the behavioural task.

Page 12-13, lines 289-327: *"In the current work, patients, most of whom had temporal lobe epilepsy, performed poorly in the explicit recognition test as patients with MTL lesions. By contrast, they presented robust neural tuning at target frequencies corresponding to different levels of the speech hierarchy (i.e., word, syllable, and pair of syllables) during the learning phase. This result indicates that learning did take place and that the hippocampus was functional with respect to statistical learning. It also confirms that implicit online measures of learning based on electrophysiological data are more sensitive than behavioural measures (François, Tillmann & Schön, 2012). Indeed, the analysis of the ERPs collected during the 2AFC task also revealed significant differences between words and nonwords over hippocampal channels. This result fits well with previous studies on speech segmentation based on SL showing functional activations of the hippocampus during speech segmentation tasks (Turk-Browne et al., 2009; Schapiro, Kustner, & Turk-Browne 2012; Schapiro et al., 2016; Barascud et al., 2016). A similar familiarity effect has been also reported when focusing on the 2AFC test (François & Schön, 2010, 2011; De Diego Balaguer et al., 2007). These studies used scalp EEG to show that healthy adults exhibited a larger negativity for unfamiliar than for newly learned. However, the percentage of correct explicit word recognition did not differ from chance level. Similar discrepancies between behavioural and neural data have been reported in previous neuroimaging studies of speech segmentation based on SL in healthy adults (François & Schön, 2010, 2011; McNealy et al., 2006; Turk-Browne et al., 2009; Sanders et al., 2002) and in patients with MTL damage (Henin et al., 2021; Schapiro et al., 2014; Covington, Brown-Schmidt & Duff, 2018). Moreover, the role of the hippocampus and MTL region during recognition memory tasks has largely been demonstrated in both healthy adults and patients with damage to the MTL (Brown & Aggleton, 2001; Düzel et al., 2001; Eldridge et al., 2000; Stark & Squire, 2000; Ranganath et al., 2004). Here, we used an implicit procedure during the learning phase and evaluated the learning using an explicit behavioural task that requires the conscious recognition of word-forms presented auditorily. While our approach has the advantage of being of a very short duration, the 2AFC task has been largely criticized for its low sensitivity due to different factors (François, Tillmann & Schön, 2012; Batterink et al., 2015; Siegelman, Bogaerts & Frost, 2017; Siegelman et al., 2018; Frost, Armstrong & Christiansen, 2019;*

*Christiansen, 2019; ). For instance, the AFC task requires participants to make an explicit judgment on two presented items without feedback, which might be particularly challenging in the case of the relatively weak memory traces created during the implicit learning phase (Schön & François, 2011; Rodriguez-Fornells et al., 2009). Moreover, the design of the AFC test trials does not allow differentiating between word recognition and nonword rejection as it is the case when using a lexical decision task (François et al., 2016; Ramos-Escobar et al., 2021). Recent studies on speech segmentation based on SL have elegantly proposed innovative designs to overcome the weaknesses associated with the use of explicit tests. Of particular relevance is the use of implicit measures such as EEG, sEEG, or Reaction-Times collected during the learning or an online test phase (see for example François et al., 2016, 2017; de Diego Balaguer et al., 2007 for the analysis of ERPs to illegal items without explicit recognition) that seem more appropriate and sensitive to fully capture implicit learning processes (Kim, Seitz, Feenstra, & Shams, 2009; Kóbor et al., 2020; Turk-Browne et al., 2005; Batterink & Paller, 2017; Siegelman, Bogaerts & Frost, 2017)."*

Concerning the possibility of splitting the data, we followed the reviewer suggestion. However, as the reviewer can see in the figure below, the effect is not clear cut, although there is a tendency for an increase at the word frequency. This is possibily due to different learning curves in the different patients that may prevent observing a clear increase. We also tried to have a more temporally resolved analysis to explore inter-individual differences, but the estimate became too noisy when using small data sets (e.g. 8 periods of 30 seconds). We eventually decided not to report this analysis in the manuscript.

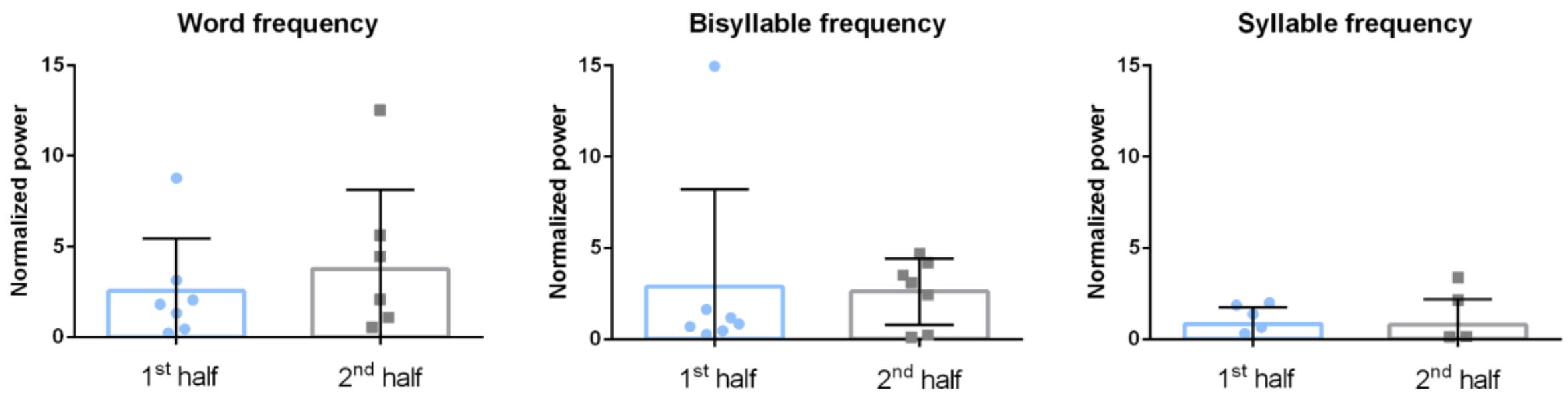

Secondary Comments

4 ms is a very short baseline period which can introduce noise to the analysis. Do the authors have justification over a longer baseline (at least 100 ms)?

Sorry this was a typo error, it should be seconds and correspond to half of the window.

The authors mention normalization in the methods. How was power normalized?
A common approach with frequency-tagging is to replot the data as signal-to-noise ratios, wherein power at the target frequency is compared against neighboring frequencies to cancel out the effects of the 1/f distribution.

We agree with the reviewer that some studies have used such a normalization procedure. However, we think that in the case of sEEG recordings the SNR is much higher than with scalp data. The suggested procedure that implicitly increases the local SNR may not be necessary in our case and we prefer not to use it and to show the 'true' FFT. Please also note that, as detailed above, we do not have the 1/f in the PSD because we work on averages. Further, recent studies have used similar approaches to study the neural mechanisms supporting the extraction of speech units based on SL in adults and children (see Jonas et al., 2016; Ordin et al., 2020; Ramos-Escobar et al., 2021).

Why was evoked power calculated as opposed to total power averaged across the entire time-range? Evoked power, when no jitter across trials, can lead to peaks at intrinsic oscillations. Moreover, total power would enable a plot of the 1/f distributions for electrodes and subjects which can be helpful in evaluating the quality of the recordings.

As we clarified above, the strategy of averaging is commonly used (see for instance *Nozaradan et al., 2021; Jonas et al., 2016*) in frequency tagging analysis to *enhance the signal-to-noise ratio of EEG activities time locked to the patterns. Below, we computed the full range power spectral density for each patient (colored lines) for both hippocampal (top) and auditory (bottom) channels. On the left, the reviewer can appreciate that it is not easy to see much on the regular PSD of hippocampal channels. The scenario becomes a little bit better when normalizing by neighbours (dividing each value by two neighbour values), as can be seen on the right part of the figure. However, while for the auditory cortex, that has a very strong response to the syllabic rate, the result is clear cut, for the hippocampal channels, have smaller responses, results are less clear and mostly visible in the first harmonic of the word frequency (2.2 Hz). We feel that this well illustrates the advantage of computing the FFT of a sliding average. Also, note that, as reported in the methods section, we cautiously use an overlap equal to twice the size of the word duration to ensure that possible artifacts would not lead to a spurious peak at the word frequency.*

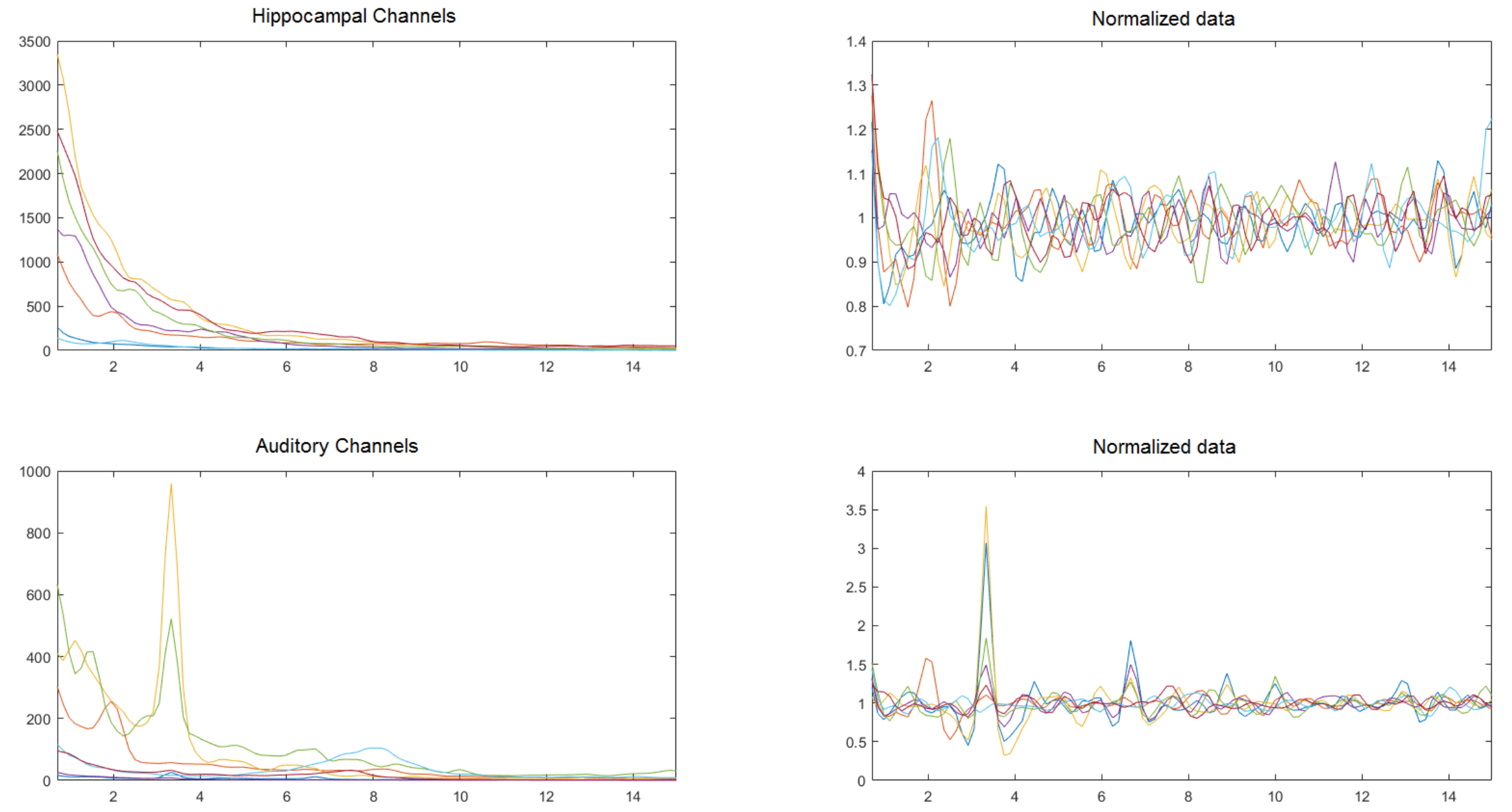

Assuming the power effects are driven by the stimuli, is it possible that the hippocampus tracked 'words' because the task required discrimination of 3 phoneme groups? Were subjects aware what they would be tested on?

We thank the reviewer for this comment. In this specific case, the answer is no. We used an implicit version of the SL paradigm in which the patients were not aware of the purpose of the task nor that they would be tested afterward. We agree that some studies have used explicit instructions of learning which may have triggered different cognitive mechanisms (Cunillera et al., 2006, 2009). Again, here, the patients were only instructed to listen carefully to an auditory stream without explicit instructions of learning. Importantly, the grouping of phonemes can only be done by statistical learning as there are no other (e.g., acoustic) cues to group the individual phonemes.

Credit Author statement.

Neus Ramos-Escobar, Manuel Mercier, Agnès Trébuchon-Fonséca, Antoni Rodriguez-Fornells, Clément François, Daniele Schön

Author contributions are reported following the CRediT Contributor Roles:

**Conceptualization:** Clément François, Daniele Schön, Antoni Rodriguez-Fornells, Agnès Trébuchon-Fonséca

**Project administration:** Daniele Schön, Clément François,

**Supervision:** Daniele Schön, Clément François

**Data curation**: Neus Ramos-Escobar, Manuel Mercier

**Writing - original draft:** Neus Ramos-Escobar, Clément François, Daniele Schön

**Writing - review & editing:** Manuel Mercier, Antoni Rodriguez-Fornells, Agnès Trébuchon-Fonséca, Clément François, Daniele Schön

# Hippocampal and auditory contributions to speech segmentation

Neus Ramos-Escobar[a,b] , Manuel Mercier[c], Agnès Trébuchon-Fonséca[c,d], Antoni Rodriguez-Fornells[a,b,e], Clément François[f]*, Daniele Schön[c]*

[a]Dept. of Cognition, Development and Educational Science, Institute of Neuroscience, University of Barcelona, L'Hospitalet de Llobregat, Barcelona, 08097, Spain.

[b]Cognition and Brain Plasticity Group, Bellvitge Biomedical Research Institute (IDIBELL), L'Hospitalet de Llobregat, Barcelona, 08097, Spain

[c]Aix Marseille Univ, Inserm, INS, Inst Neurosci Syst, Marseille, France

[d]APHM, Hôpital de la Timone, Service de Neurophysiologie Clinique, Marseille, France

[e]Catalan Institution for Research and Advanced Studies, ICREA, Barcelona, Spain

[f]Aix Marseille Univ, CNRS, LPL, (13100) Aix-en-Provence, France

* co-senior authorship

*Corresponding Authors: Daniele Schön: daniele.schon@univ-amu.fr +33 491324100 and Clément François: clement.francois@univ-amu.fr +33 413552714

**Abstract**

Statistical learning has been proposed as a mechanism to structure and segment the continuous flow of information in several sensory modalities. Previous studies proposed that the medial temporal lobe, and in particular the hippocampus, may be crucial to parse the stream in the visual modality. However, the involvement of the hippocampus in auditory statistical learning, and specifically in speech segmentation is less clear. To explore the role of the hippocampus in speech segmentation based on statistical learning, we exposed seven pharmaco-resistant temporal lobe epilepsy patients to a continuous stream of trisyllabic pseudowords and recorded intracranial stereotaxic electro-encephalography (sEEG). We used frequency-tagging analysis to quantify neuronal synchronization of the hippocampus and auditory regions to the temporal structure of words and syllables of the learning stream. We also analyzed the event-related potentials (ERPs) of the test to evaluate the role of both regions in the recognition of newly segmented words. Results show that while auditory regions highly respond to syllable frequency, the hippocampus responds mostly to word frequency. Moreover, ERPs collected in the hippocampus show clear sensitivity to the familiarity of the items. These findings provide direct evidence of the involvement of the hippocampus in the speech segmentation process and suggest a hierarchical organization of auditory information during speech processing.

**Keywords:** Hippocampus, statistical learning, frequency tagging, SEEG, speech segmentation

## Introduction

Humans are daily exposed to a massive amount of information. Finding a structure in the sensory flow is necessary to make sense of the world. A structure can emerge thanks to regularities in the input tracked by computing low-order statistics (Reber, 1967; Frost et al., 2015). Statistical learning (SL) is a domain-general learning mechanism through which learners track statistical regularities of motor (Hunt & Aslin, 2001), visual (Fisher & Aslin, 2002), and auditory sequences (Saffran et al., 1996, 1999; see Frost et al., 2015 for a review).

Speech segmentation is one of the first problems that language learners must deal with when learning a new language (Graf-Estes et al., 2007; François et al., 2017). SL has been proposed as a possible mechanism that allows segmenting words from fluent speech (Cutler & Butterfield, 1992; Saffran et al., 1996). This process can occur incidentally and without effort via simple exposure, as in the case of infants (Saffran et al., 1997; Turk-Browne et al., 2005; Saffran et al., 1999). Although several behavioral (Cutler & Butterfield, 1992; Saffran et al., 1996; Schön et al., 2008) and electrophysiological studies (Sanders et al., 2002; Cunillera et al., 2006; de Diego-Balaguer et al., 2007; Abla et al., 2008; François et al., 2014; 2017) have explored the bases of SL, the underlying precise brain network dynamics are not clear yet.

Capitalizing on a high spatial resolution, functional magnetic resonance imaging (fMRI) studies have allowed to decipher the brain regions supporting SL in the auditory and visual modalities. Results showed activations of modality-specific brain regions during exposure to learning streams (Turk-Browne et al., 2009; Bischoff-Grethe et al., 2000; McNealy et al., 2006; Cunillera et al., 2009; Karuza et al., 2013). Specifically, fMRI speech segmentation studies consistently observed functional activations of typical language areas such as the middle and superior temporal regions (MTG & STG) and the inferior frontal gyrus (IFG; McNealy et al., 2006; Cunillera et al., 2009; Karuza et al., 2013). However, activations of the hippocampus were also observed in a few SL studies (Turk-Browne et al., 2009; Schapiro, Kustner, & Turk-Browne 2012; Schapiro et al., 2016; Barascud et al., 2016). The interplay between cortical and subcortical structures during SL fits well with cognitive models proposing that complementary neural systems may account for human learning abilities (Davis & Gaskell, 2009; McClelland et al., 1995). Specifically, these models suggest that learning and memory processes may occur in two different stages. The medial temporal structures would support the initial acquisition and formation of memory traces, while neocortical regions may participate in their long-term storage. Interestingly, the hippocampus has been proposed to play a crucial role in segmenting continuous sensory inputs into discrete events (Radvansky & Zacks, 2017). Recent studies on event memory formation propose that the interplay between sensory regions and the hippocampus may support the creation of boundaries between events. Specifically, while sensory areas seem to be responsible for fine-grained boundaries, the hippocampus instead supports cortical information binding into memory traces (Baldassano et al., 2017; Ben-Yakov & Dudai, 2011; Zacks et al., 2001; Speer et al., 2007). Further, recent studies on vocabulary acquisition based on associative or contextual learning consistently show functional activations of the hippocampus during the early stages of learning (Bartolotti et al., 2017; Breitenstein et al., 2005; Covington & Duff, 2016; Ripollés et al., 2016; Züst et al., 2019). However, direct human electrophysiological evidence for the role of the hippocampus in extracting pattern regularities in speech is still missing.

Recently, electrophysiological studies have capitalized on the brain property to oscillate at the frequency of a continuous auditory stimulus to explore the neural mechanisms supporting the hierarchical processing of speech and music (Nozaradan et al., 2014; Giraud & Poeppel, 2012; Poeppel & Teng, 2020). Specifically, frequency tagging analysis have been successfully applied to surface EEG or MEG recordings to quantify the amount of neural synchronization to syllable, pairs of syllables and words during speech segmentation tasks (Buiatti et al., 2009; Ding et al., 2016; Batterink & Paller, 2017). In a recent study, Henin and colleagues (2020) collected intracortical brain responses from human epileptic patients during an auditory and a visual SL task. They applied frequency-tagging to electrocorticography (EcoG) data to show that neural response in the STG synchronized to both syllables and word frequency. They also found synchronized neural response to word frequency in the IFG and Anterior Temporal Lobe. However, no evidence of neural synchronization was observed in the hippocampus possibly due to a limited access provided by EcoG probes. Nonetheless, using a more indirect method based on multivariate pattern similarity analysis, they were able to show the involvement of the hippocampus in word identity during learning.

Here, we gathered intracranial recordings from 7 patients with pharmaco-resistant temporal lobe epilepsy implanted with depth electrodes to directly assess the contribution of the auditory cortex and the hippocampus during a speech segmentation task based on SL. Participants passively listened to 4 minutes of an artificial statistically structured speech stream and were tested on their ability to recognize the newly segmented words. We used frequency-tagging to quantify the level of neural synchronization in auditory and hippocampal regions to the constitutive elements of the inputs, namely syllables, pairs of syllables and tri-syllabic words during the learning phase. We expected auditory regions to show a peak in the power spectrum corresponding to the syllable rate reflecting phonological processing, while the hippocampus was expected to exhibit high neural synchronization to pairs of syllables and word frequencies, reflecting its role in speech segmentation. Moreover, previous reports studying memory have extensively shown the involvement of the hippocampus (Ripollés et al., 2016;

Brown & Aggleton, 2001; Düzel et al., 2001; Eldridge et al., 2000; Stark & Squire, 2000; Ranganath et al., 2004). Therefore, we also analyzed the event-related potentials (ERPs) collected during the behavioural test to evaluate the contribution of both regions during the recall of newly segmented words.

## Methods

### Participants

Seven patients with pharmaco-resistant temporal lobe epilepsy (4 females, mean age = 29; range 18-45) participated in the study (see **Table 1**). Patients were implanted with depth electrodes for clinical reasons to determine the epileptic zone before they underwent neurosurgical treatment at the La Timone Hospital in Marseille (France). The location of the implanted electrodes was solely determined by clinical criteria. Patients provided informed consent prior to the experimental session, and the study was approved by the Institutional Review Board of the French Institute of Health (IRB00003888). No part of the study procedures was pre-registered prior to the research being conducted.

**Table 1:** Patients clinical description

| Patients | Gender | Age (years) | Hemispheric dominance | Epileptogenic zone | Depth electrodes | Hippocampal electrodes |
|---|---|---|---|---|---|---|
| P1 | F | 29 | L | L temporal | 4R + 10L | Both |
| P2 | F | 45 | L | R temporal | 10R + 2L | Both |
| P3 | F | 18 | L | R temporal | 5R + 4L | Both |
| P4 | F | 23 | Atypical | L temporal | 1R + 12L | L |
| P5 | M | 19 | L | L temporal | 2R + 11L | R |
| P6 | M | 42 | L | L Frontal | 1R + 13L | L |
| P7 | M | 33 | L | R Frontal & Parietal | 14R | R |

*M* male, *F* female, *L* left, *R* right

### Data acquisition & electrode localization

The sEEG signal was recorded using depth electrodes of 0.8 mm diameter containing 10 to 15 electrodes contacts (Alcis, Besançon, France). The electrode contacts were 2 mm long and were spaced from each other by 1.5 mm. Data was recorded using a BrainAmp amplifier system (Brain Products GmbH, Munich, Germany), sampled at 1000 Hz and high-passed filtered at 0.016 Hz. During the acquisition, recordings were referenced to a single scalp-electrode located at Cz. Contact data was offline converted to virtual channels using a bipolar montage approach (closest-neighbor contact reference) to increase spatial resolution and reduce passive volume diffusion from neighboring areas (Mercier et al., 2017).

To precisely localize the channels, a procedure similar to the one used in the iELVis toolbox was applied (Groppe et al., 2017). First, we manually identified the location of each channel centroid on the post-implant CT scan using the Gardel software (Medina et al., 2018). Second, we performed volumetric segmentation and cortical reconstruction on the pre-implant MRI with the Freesurfer image analysis suite (documented and freely available for download online http://surfer.nmr.mgh.harvard.edu/). Third, we mapped channel locations to the pre-implant MRI brain (processed with FreeSurfer) and to the MNI template, using SPM12 methods (Penny et al., 2011), through the FieldTrip toolbox (Oostenveld et al., 2011). The co-registration to the patient brain was done via a rigid, affine transformation to respect individual anatomy. The normalization to the MNI template was done through a non-linear transformation to map channels to a standardized space and allow brain regions labeling using the Destrieux atlas (Destrieux et al., 2010). The definition of hippocampal and primary auditory channels was determined using a combination of automatic atlas labeling and visual inspection of the anatomical data in 2D and 3D representations (see **Figure 1**).

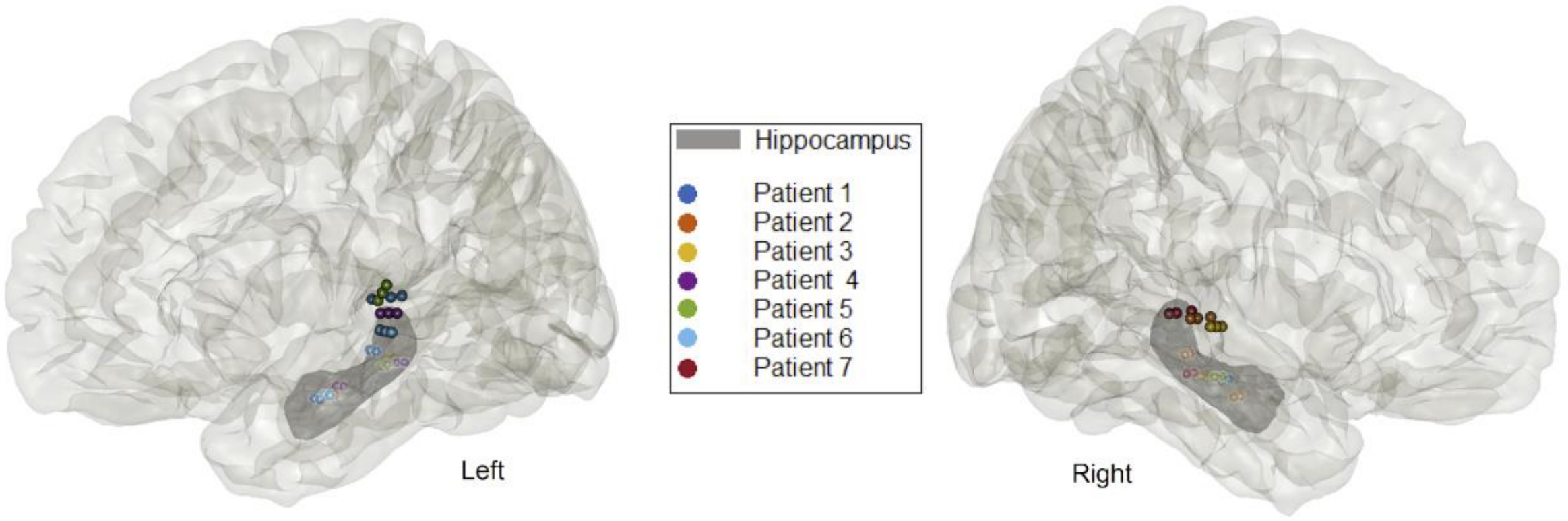

**Figure 1.** sEEG channel location. Colored dots indicate the channel location for each patient in auditory (dark-colored) and hippocampal (light-colored) regions. Light gray represents the cortical sheet of the FreeSurfer brain template. The shaded area depicts the hippocampus.

## Experimental procedure

We used a similar experimental design to the one used in our previous studies with healthy adults and children (Schön et al., 2008; François & Schön 2010; 2011; François et al., 2013; 2014). Specifically, the experimental procedure consisted of two consecutive phases, an implicit learning phase followed by an explicit 2-alternative forced-choice (2AFC) task. Before starting the implicit learning phase, patients were asked to listen carefully to one single auditory stream without explicit instructions of learning (see Stimuli section for a description of the speech streams). Importantly, we did our best to keep the entire procedure implicit. During the learning phase, patients were exposed to a single continuous speech stream that was composed of 4 pseudo-words presented 60 times each, thus leading to a single continuous stream of 240 words that lasted 4 min. Immediately after this learning phase, patients performed the behavioural 2AFC task that lasted 5 min. During each trial of the test, patients were presented with two consecutive auditory words and had to press one of two buttons to indicate which of two words (first or second item) most closely resembled what they had just heard in the continuous stream (see Figure 2). Importantly, one test item was a word from the learning stream while the other was a "nonword" that was never heard before the test. Each familiar word of the language (word) was presented with each unfamiliar word (nonwords), making up 16 pairs that were repeated twice, thus leading to 32 test trials.

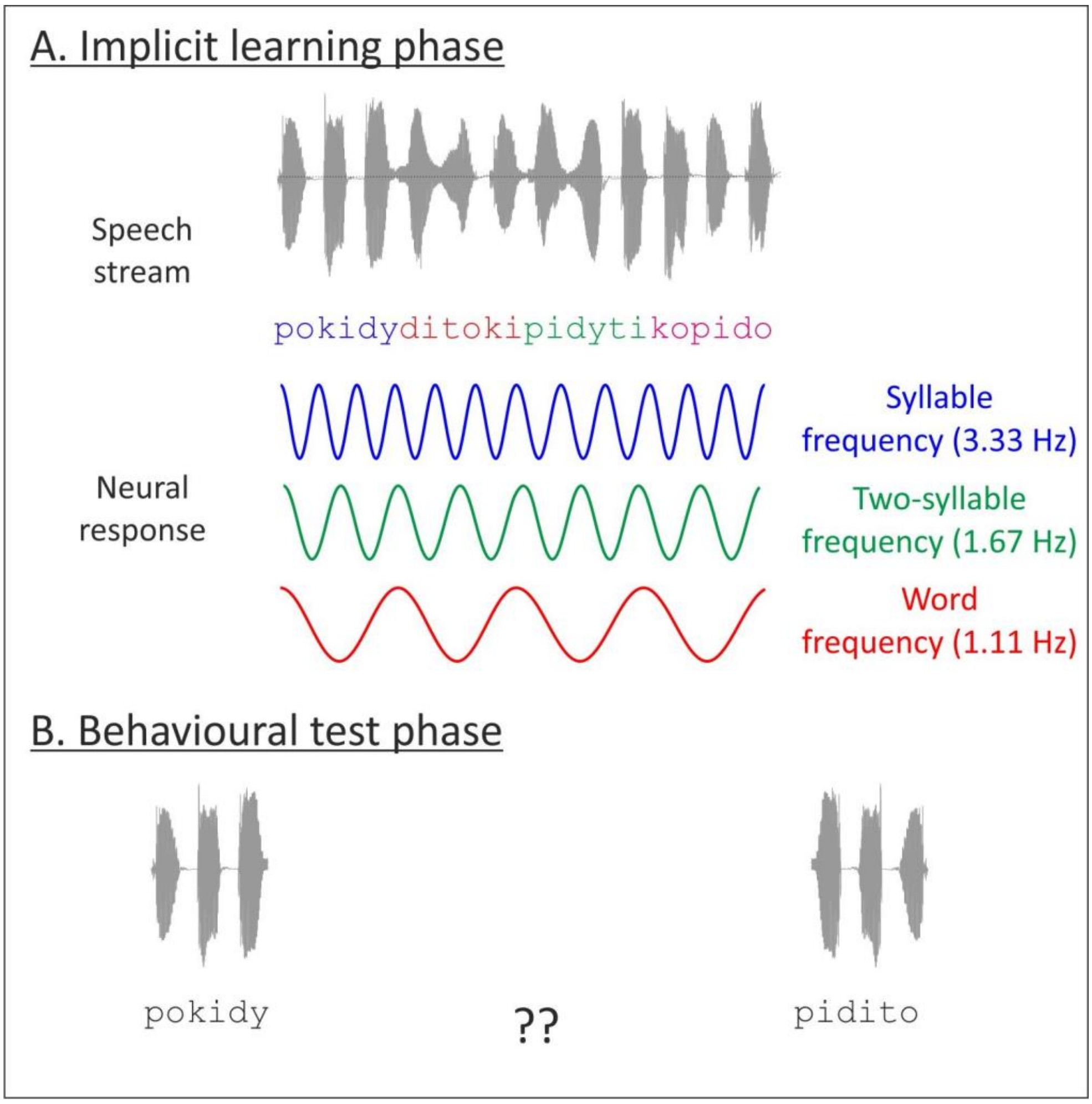

**Figure 2.** Illustration of the experimental procedure. After being exposed to a continuous stream of statistically structured syllables/words without instruction of learning (A), participants performed a 2AFC task to assess the level of learning (B). The auditory cortex should preferentially respond to the syllable frequency reflecting the tracking of low-order speech structure.

The hippocampus should preferentially respond to the word frequency reflecting the creation of event boundaries during the learning.

## Stimuli

The language consisted of four consonants (‘p’, ‘t’, ‘k’, ‘d’) and three vowels (‘o’, ‘i’, ‘y’), which were combined into a set of eleven syllables. The exact syllable length was set to 300 ms. These syllables were then combined to give rise to 4 tri-syllabic words (POKIDY, DITOKI, PIDYTI, and KOPIDO). The stream was built by random concatenation of the four pseudowords and synthesized using Mbrola (http://tcts.fpms.ac.be/synthesis/mbrola.html). More precisely, the speech stream was built by concatenating seven minimal sequences of non-coarticulated syllables respecting the constraint of not repeating the same word twice in a row. Importantly, no acoustic cues have been inserted at word boundaries. In the test, the items consisted of the four words used in the learning phase and four nonwords created by pseudo-randomly mixing the syllables of the words from the language TOPIDY, DYPOKI, KOKITI, and PIDITO.

## SEEG Data analysis: Frequency tagging (learning phase)

For each patient, sEEG data, in a bipolar montage, were visually inspected using AnyWave software (Colombet et al., 2015), and channels with artifacts or epileptic activity were excluded from the analysis. Continuous sEEG recordings acquired during the learning task were filtered using a 0.5 Hz high pass filter to remove slow drifts in the recorded signal. Then, epochs time-locked to the onset of each word were created by segmenting the continuous EEG data from 4 words before and 4 after the stimulus yielding epochs of 8-word length (lasting 7.2 s). Epochs were partially overlapping, yet we took care to use an overlap equal to twice the size of the word to ensure that possible artifacts would not lead to a spurious peak at the word frequency. A baseline correction was applied (-3.6 to 0 s). Epochs with high amplitude values were excluded (threshold: mean +2 SD). Epochs were averaged and transformed to the frequency domain using a discrete Fourier transformation (Matlab; Natick, MA). Importantly, by computing averages, similarly to other frequency tagging studies (Nozaradan et al., 2021; Jonas et al., 2016), we remove non time-locked activity (intrinsic oscillations), enhance the signal-to-noise ratio of EEG activities time locked to the patterns and only focus on evoked activity. We extracted the power values for each target frequency (word frequency: 1.11 Hz; two-syllables frequency: 1.67 Hz; syllable frequency: 3.33 Hz). Power values at the target frequencies were obtained for each patient and channel.

## SEEG Data analysis: ERP analysis (Test phase)

We used a similar strategy with the sEEG data collected during the 2AFC test. First, we changed to a bipolar montage to increase spatial resolution, high-pass filtered at 0.5 Hz and low-pass filtered at 20

Hz. Then, we created epochs time-locked to the item onset using a -100 ms 1200 ms time-window. A baseline correction was applied (-100 to 0 ms). We only report analyses of channels in the hippocampus and the primary auditory cortex.

## Statistical analyses

For each patient and for each target frequency (word, syllable & two syllables), we computed the distribution of power values across all contacts (between 140 and 200 contacts per patient, spanning several brain regions beyond the primary auditory cortex and the hippocampus). Since the distribution was not normal, we used a non-parametric threshold (median + 2.5 interquartile range, IQR) to determine whether hippocampal and auditory contacts showed a significant response at the target frequencies, as compared to overall channels (see **Figure 3**).

Whenever more than one channel was present in the same region (primary auditory or hippocampus), the average power values of the two channels was used. For patients with bilateral implantation and artifact free hippocampi, the average power values of channels located in both hemispheres was used. Finally, to assess the power differences between hippocampal and auditory channels for each patient at word, two-syllable, and syllable frequencies, we normalized the data across channels for each frequency and patient and applied the Wilcoxon test.

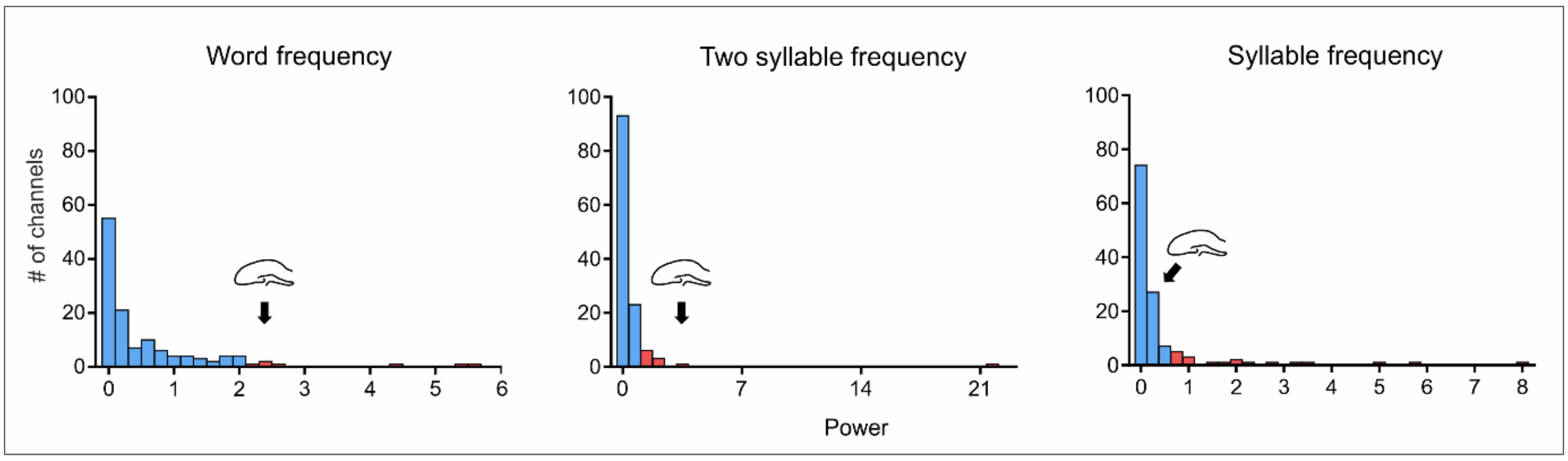

**Figure 3.** Example of the methodology used to define significant hippocampal implication. Histograms of power response of all contacts (N ~ 150) to word, two-syllable, and syllable target frequencies for Patient 6. Power values above the threshold (median plus 2.5 IQR) are represented by red bars. Black arrows indicate the frequency bins where the hippocampal power response falls. In this example, the hippocampal response is significant at the word and two-syllable frequencies (arrow on red bars) but not at the syllable frequency (arrow on blue bars).

To analyze the ERP data of the test phase, we first compared the amplitude of the ERPs to words and nonwords using mean amplitude values in successive 50 ms time-windows between 250 and 700 ms post-stimulus onset. Then, we computed a mixed-model including each trial (one value per trial per condition per patient: val~conditions+trials+(1|subjects)).

## Results

*Test phase*: The level of performance in the 2AFC test reveals that the percentage of correct explicit word recognition did not differ from chance level (range: 25-56%, $p > .05$, wilcoxon signed-rank) thus confirming previous results of impaired explicit word recall in patients with epilepsy (Schapiro et al., 2014; Henin et al., 2021). Importantly, however, as shown on **Figure 4**, the ERP data show a significant difference between words and nonwords in hippocampal channels in the 250-400 (beta = -18.8; CI = -33.3 -4.2; $p$ <.01) and 550-700 ms (beta = -19.6, CI = -35.9 -3.2; $p < .01$) time-windows. A significant effect over a single 50 ms time window, between 350 and 400 ms, is also found over auditory channels (beta = -8.4, CI = -16.5 -0.7; $p < .05$). Overall, these results confirm that patients did segment the words during the learning phase and that the hippocampus is particularly sensitive to the familiarity of the items.

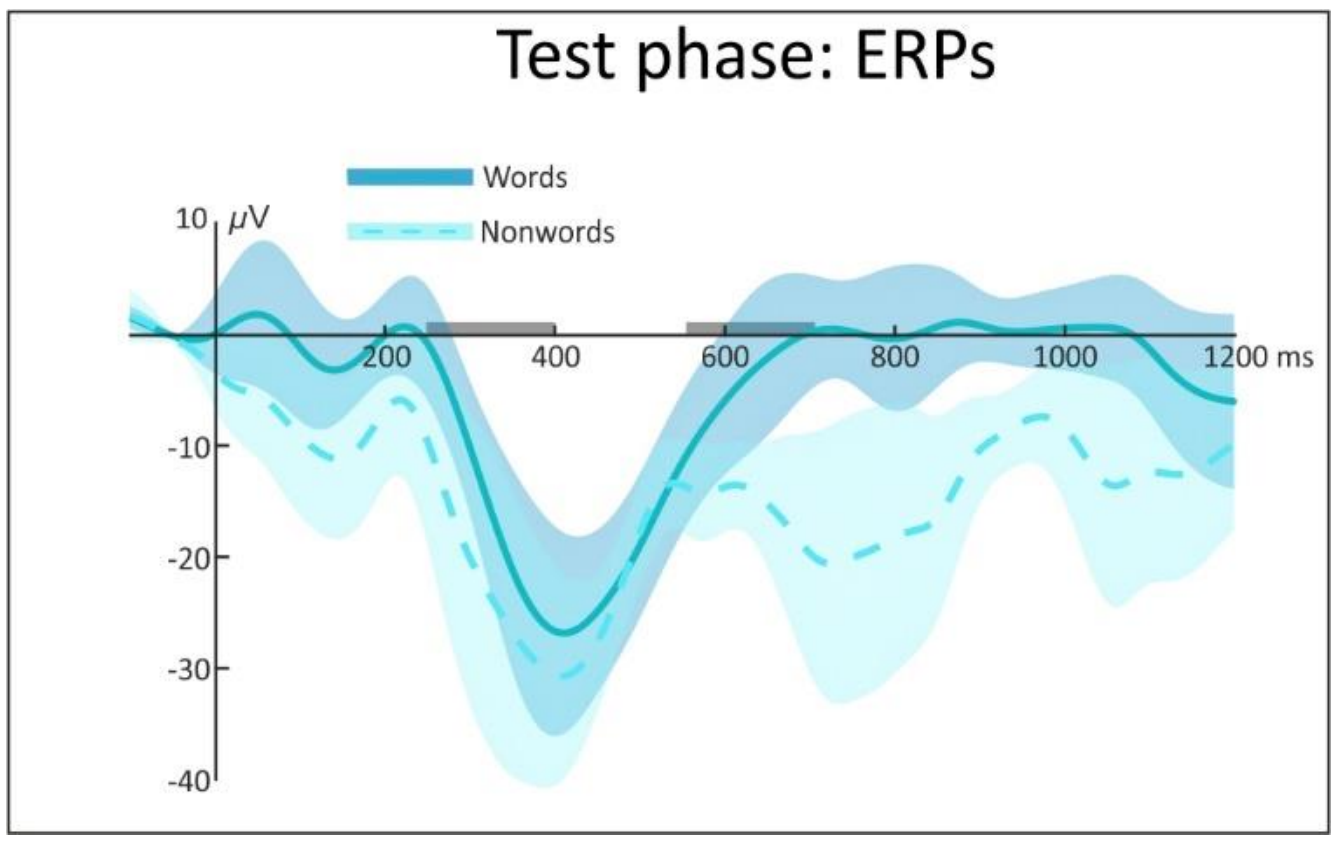

**Figure 4.** ERPs to words and nonwords in hippocampal contacts (bipolar montage) averaged across 6 patients obtained during the 2AFC task. The thick and dashed lines show the mean of ERPs to words and nonwords respectively. The shaded areas correspond to the standard error of the mean in each condition. The grey areas depict the two time-windows showing significant differences between the two conditions.

*Learning phase*: Clear power spectrum peaks at word and syllable frequencies are visible over auditory and hippocampal contacts (see **Figure 5A**).

For the syllable frequency, all patients except one exhibited a clear peak in contacts located within the primary auditory cortex (raw data median = 12.24; IQR = 315.69). Five patients also showed significant responses at this target frequency in hippocampal contacts although much smaller than auditory responses (raw data median = 1.62; IQR = 2.76).

For the word-frequency, all patients except one (Patient 4) showed a significant response in hippocampal contacts (raw data median = 3.86; IQR = 15.95). Three patients also showed a significant response to word-frequency in auditory contacts although smaller than hippocampal responses (raw data median = 1.62; IQR = 8.73).

For the two-syllable frequency, all patients showed a significant response at hippocampal contacts (raw data median = 4.79; IQR = 5.87). By contrast, none of the patients showed a significant response to the two-syllable frequency in auditory contacts (raw data median = 0.59; IQR = 0.71).

The amplitude of the peaks in the power spectrum of the hippocampus differed from that in auditory regions across all target frequencies (word frequency: Cohen d = 0.5; *p* = .01; two-syllable frequency: d = 0.46; *p* = .01; syllable frequency: d = 0.7; *p* = .03).

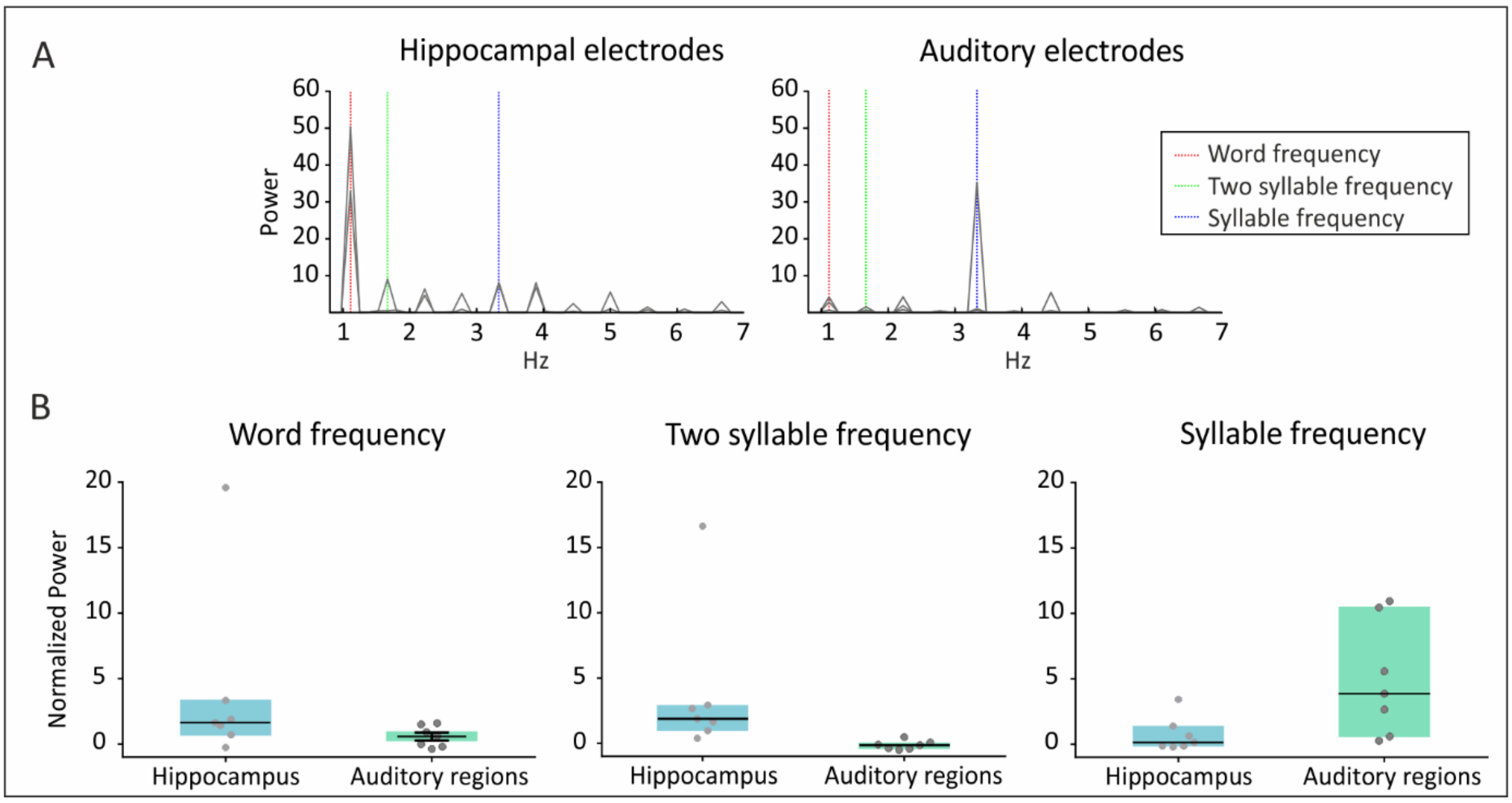

**Figure 5.** A) Example of a patient (Patient 7) power response of hippocampal and auditory electrodes to word frequency (red), two-syllable frequency (green) and syllable frequency (blue). B) Average of all patients' neural responses to word, two-syllables and syllable frequencies in hippocampus and auditory regions (z-score normalized data). Black lines indicate the median of all patients and box plots indicate the interquartile range.

## Discussion

In the present study, we directly assessed the contribution of auditory regions and the hippocampus during speech segmentation based on SL. Pharmaco-resistant epileptic patients implanted with sEEG depth electrodes listened to a continuous stream of statistically organized syllables. The frequency-tagging analysis reveals that the hippocampus preferentially responds to word-frequency. By contrast, auditory regions preferentially tune their response to syllable frequency (see **Figure 5B**). Although previous studies have suggested the involvement of MTL regions and especially the hippocampus in SL based on indirect measures, we provide the first direct evidence for its role during speech segmentation based on SL.

Previous neuropsychological studies showed that patients with lesions of the MTL are impaired in extracting auditory and visual statistical patterns (Schapiro et al., 2014; Covington, Brown-Schmidt

& Duff, 2018). In a single case study, a patient with complete bilateral loss of hippocampus and extensive damage to surrounding MTL regions could not recall familiar sequences in a visual SL task (Schapiro et al., 2014). However, Covington and colleagues (2018) showed that patients with hippocampal damage could perform above chance level in SL tasks, although they were overall impaired in comparison to healthy controls. Therefore, although the hippocampus might participate and to a certain extent facilitate statistical learning by strengthening associations between input elements, its participation might not be strictly necessary and other non-hippocampal cortical regions could support SL.

In the current work, patients, most of whom had temporal lobe epilepsy, performed poorly in the explicit recognition test as patients with MTL lesions. By contrast, they presented robust neural tuning at target frequencies corresponding to different levels of the speech hierarchy (i.e., word, syllable, and pair of syllables) during the learning phase. This result indicates that learning did take place and that the hippocampus was functional with respect to statistical learning. It also confirms that implicit online measures of learning based on electrophysiological data are more sensitive than behavioural measures (François, Tillmann & Schön, 2012). Indeed, the analysis of the ERPs collected during the 2AFC task also revealed significant differences between words and nonwords over hippocampal channels. This result fits well with previous studies on speech segmentation based on SL showing functional activations of the hippocampus during speech segmentation tasks (Turk-Browne et al., 2009; Schapiro, Kustner, & Turk-Browne 2012; Schapiro et al., 2016; Barascud et al., 2016). A similar familiarity effect has been also reported when focusing on the 2AFC test (François & Schön, 2010, 2011; De Diego Balaguer et al., 2007). These studies used scalp EEG to show that healthy adults exhibited a larger negativity for unfamiliar than for newly learned. However, the percentage of correct explicit word recognition did not differ from chance level. Similar discrepancies between behavioural and neural data have been reported in previous neuroimaging studies of speech segmentation based on SL in healthy adults (François & Schön, 2010, 2011; McNealy et al., 2006; Turk-Browne et al., 2009; Sanders et al., 2002) and in patients with MTL damage (Henin et al., 2021; Schapiro et al., 2014; Covington, Brown-Schmidt & Duff, 2018). Moreover, the role of the hippocampus and MTL region during recognition memory tasks has largely been demonstrated in both healthy adults and patients with damage to the MTL (Brown & Aggleton, 2001; Düzel et al., 2001; Eldridge et al., 2000; Stark & Squire, 2000; Ranganath et al., 2004). Here, we used an implicit procedure during the learning phase and evaluated the learning using an explicit behavioural task that requires the conscious recognition of word-forms presented auditorily. While our approach has the advantage of being of a very short duration, the 2AFC task has been largely criticized for its low sensitivity due to different factors (François, Tillmann & Schön, 2012; Batterink et al., 2015; Siegelman, Bogaerts & Frost, 2017; Siegelman et al., 2018; Frost, Armstrong & Christiansen, 2019; Christiansen, 2019; ). For instance, the AFC task requires participants to make an explicit judgment on two presented items without feedback, which might be particularly challenging in the case of the relatively weak memory traces created during the implicit learning phase (Schön & François, 2011; Rodriguez-Fornells et al., 2009). Moreover, the design of the AFC test trials does not allow differentiating between word recognition and nonword rejection as it is the case when using a lexical decision task (François et al., 2016; Ramos-Escobar et al., 2021). Recent studies on speech segmentation based on SL have elegantly proposed innovative designs to overcome the weaknesses associated with the use of explicit tests. Of particular relevance is the use of implicit measures such as EEG, sEEG, or Reaction-Times collected during the learning or an online test phase (see for example François et al., 2016, 2017; de Diego Balaguer et al., 2007 for the analysis of ERPs to illegal items without explicit recognition) that seem more appropriate and sensitive to fully capture implicit learning processes (Kim, Seitz, Feenstra, & Shams, 2009; Kóbor et al., 2020; Turk-Browne et al., 2005; Batterink & Paller, 2017; Siegelman, Bogaerts & Frost, 2017).

Previous studies with surface EEG or MEG have successfully used frequency tagging to track the patterns of cortical synchronization supporting the hierarchical processing of speech (Buiatti et al., 2009; Ding et al., 2016; Batterink & Paller., 2017; see Poeppel & Teng, 2020 for a review). Importantly however, while functional activations of the hippocampus have been consistently reported during visual SL tasks (Turk-Browne et al., 2009; Schapiro, Kustner & Turk-Browne 2012), this was not the case using sequences of syllables (McNealy et al., 2006; Cunillera et al., 2009; Karuza et al., 2013). Further, in a recent study, Henin and colleagues gathered brain responses to statistically structured auditory and visual sequences in 26 patients with MTL epilepsy (Henin et al., 2021). Using similar frequency tagging analysis applied to EcoG data, they found clear neural response at both two-syllable and word frequencies over multiple cortical regions. However, evidence for a contribution of the hippocampus was only observed with a more indirect analysis based on representational similarities (dissimilarity measures). Here, instead of using grid electrodes located at the surface of the cortex (referenced to subdural/skull contacts), we used depth sEEG electrodes and in particular bipolar montages that allow a high spatial resolution and directly quantifying neural response at the population level in the auditory cortex and in the hippocampus. Results are clear cut in showing that auditory regions significantly respond to syllable frequency but not to word frequency. Crucially, we observe an opposite pattern in the hippocampus with an ample response to longer units (i.e., pairs of syllables and words, see **Figure 5B**).

These results strongly corroborate a hierarchical organization of auditory information during speech segmentation. Moreover, the hippocampal response to both pairs of syllables and word frequencies sheds light on the neural validity of speech segmentation models. According to the PARSER model, continuous speech is segmented by extracting small chunks of increasing size based on the computation of temporal proximity and associative learning mechanisms. Through repetition, these chunks are consolidated and stored, allowing explicit behavioural recognition of the newly learned items (Perruchet & Vinter, 1998). More recent work on event memory formation for spatial or temporal sequences proposes that sensory regions and the hippocampus hierarchically contribute to creating boundaries between events contained in long passages (Baldassano et al., 2017; Radvansky & Zacks, 2017; Ben-Yakov & Dudai, 2011; see also Zacks & Swallow, 2007). For instance, the encoding and recall of narratives may involve the encoding of small temporal chunks in primary sensory regions. Long events encoding would occur in higher-level brain regions, including cortical areas and the hippocampus (Baldassano et al., 2017). Importantly, Schapiro and colleagues (2017) recently proposed a neuroanatomically plausible model of hippocampal functioning during continuous sequence learning such as SL. Specifically, they exposed an artificial neural network mimicking the functional and anatomical properties of the hippocampus to continuous sequences of items with different temporal regularities. Results suggested the existence of complementary learning systems in the hippocampus where specific neural pathways differently contribute to learning depending on the type of input. Our findings are in line with the idea that the hippocampus is sensitive to pattern regularities found in the environment. It seems reasonable to think that the hippocampus is also sensitive to the co-occurrence of syllable pairs as for visual sequences (Schapiro et al., 2017; Turk-Browne et al., 2009). Taken together, our data suggest a hierarchical organization of auditory information during speech processing, where both cortical and hippocampal regions contribute to language learning. While the clear response at syllable frequency in primary auditory areas may reflect the tracking of the phonological structure, the hippocampus would be involved in the encoding and storage of larger units as previously proposed in different neurocomputational models of chunking (Baldassano et al., 2017; Schapiro et al., 2017). Taken together, our data suggest that the hippocampus plays an important role in speech segmentation and language learning using a more direct measure of neural activity than previously described (Schapiro et al., 2014; Covington, Brown-Schmidt & Duff, 2018; Duff & Brown-Schmidt, 2012; Kepinska et al., 2018).

Nonetheless, our study presents methodological limitations that prevent us from drawing definite conclusions on the role of the hippocampus in speech segmentation in the general population. First, the complex clinical history of these temporal lobe epileptic patients may affect verbal memory storage and executive functions thus, explaining impaired performance at test (Zamarian et al., 2011; Saling, 2009; Squire et al., 2004). Second, while there is evidence for left lateralized activations in the Inferior and Superior Temporal Gyri during speech segmentation based on SL (Cunillera et al., 2009; McNealy et al., 2006; Karuza et al., 2013), it is still unclear as to whether asymmetric processing also takes place in the hippocampus. In our small population, only one of the patients (P4), implanted over the left hemisphere, did not significantly respond to word frequency in the hippocampus. Clinical exploration revealed that this patient had an atypical language dominance to the right hemisphere, probably induced by a disease-related atypical functioning of the hippocampus. Thus, further work on a larger sample and possibly bilateral implantations is needed to explore the possibility of a hippocampal asymmetry. Finally, Schapiro and colleagues (2017) showed that the anterior part of the hippocampus where the monosynaptic pathway connects the entorhinal cortex to the "*cornu ammonis 1*" is more involved in SL than the posterior part. Again, determining possible functional differences related to topographical gradients in hippocampal structures will require further investigations with a larger number of patients.

## Conclusion

Here, we directly assessed the role of the hippocampus in speech segmentation based on SL. We showed that the hippocampus neural response synchronizes with the word-level time scale but not with the syllable-level time scale. Conversely, auditory regions consistently responded to syllable frequency but not to word frequency. Moreover, we found clear neural evidence for the contribution of the hippocampus in the recall of newly segmented words. These findings provide preliminary but direct evidence in humans for the involvement of the hippocampus in the brain network that orchestrates auditory speech segmentation based on SL.

**Acknowledgments:** We kindly thank all the patients and their families that participated in the study. We also want to thank Patrick Marquis for his help and collaboration in the project. This research was supported by grants ANR-16-CE28-0012-01 (RALP), ANR-16-CONV-0002 (ILCB), and the Excellence Initiative of Aix-Marseille University (A*MIDEX).

**Financial Disclosures:** All the authors report no biomedical financial interests or potential conflicts of interest.